\documentclass[superscriptaddress,twocolumn,prl]{revtex4}

\usepackage{graphicx}
\usepackage{multirow}
\usepackage{bm}
\usepackage{times}
\usepackage{float}

\usepackage{color}
\usepackage{amstext}
\usepackage{amsmath}
\usepackage{amsfonts}

\begin{document}

\title{Exciton doublet in the Mott-Hubbard LiCuVO$_4$ insulator 
identified by spectral ellipsometry}
\author{Y. Matiks}
\affiliation{Max-Planck-Institut f\"{u}r Festk\"{o}rperforschung,
Heisenbergstrasse 1, D-70569 Stuttgart, Germany}
\author{P. Horsch}
\affiliation{Max-Planck-Institut f\"{u}r Festk\"{o}rperforschung,
Heisenbergstrasse 1, D-70569 Stuttgart, Germany}
\author{R.K. Kremer}
\affiliation{Max-Planck-Institut f\"{u}r Festk\"{o}rperforschung,
Heisenbergstrasse 1, D-70569 Stuttgart, Germany}
\author{B. Keimer}
\affiliation{Max-Planck-Institut f\"{u}r Festk\"{o}rperforschung, Heisenbergstrasse 1, D-70569 Stuttgart, Germany}
\author{A.V. Boris}
\affiliation{Max-Planck-Institut f\"{u}r Festk\"{o}rperforschung, Heisenbergstrasse 1, D-70569 Stuttgart, Germany}
\affiliation{Department of Physics, Loughborough University, Loughborough,
LE11 3TU, United Kingdom}
\date{\today}
\begin{abstract}
Spectroscopic ellipsometry was used to study the dielectric function of LiCuVO$_{4}$, a compound comprised of chains of edge-sharing CuO$_4$ plaquettes, in the spectral range \(0.75 - 6.5\) eV at temperatures \(7-300\)
K. For photon polarization along the chains, the data reveal a
weak but well-resolved two-peak structure centered at 2.15 and 2.95 eV whose spectral weight is strongly enhanced upon cooling near
the magnetic ordering temperature. 
We identify these features as an exciton doublet
in the Mott-Hubbard gap that emerges as a consequence of the Coulomb interaction between electrons on nearest and next-nearest neighbor sites along the chains.
Our results and methodology can be used to address the role of the long-range Coulomb repulsion for compounds with doped copper-oxide chains and planes.
\end{abstract}

\pacs{78.20.-e, 71.27.+a, 71.35.-y, 78.67.-n}

\maketitle

The elementary charge excitations observed by optical spectroscopy are a key source of information on electronic correlations in transition metal oxides. Because of their particularly simple electronic structure (with a single hole in the $d$-electron shell) and their importance for high-temperature superconductivity and other electronic ordering phenomena, insulating copper oxides have long been recognized as model systems for charge excitations in the strong-correlation limit. In these systems, the electron-hole pairs created by optical excitation, $d^9_{i\uparrow} d^9_{j\downarrow} \rightarrow (d^9L_h)_i d^{10}_j$, form the Zhang-Rice-singlet (ZRS) state \cite{Zha88} with an oxygen-ligand hole in one CuO$_4$ plaquette and the full-shell $d^{10}_j$ configuration in a neighboring plaquette. In formal analogy to the single-band Hubbard model \cite{Toh04,Ste90,Miz98}, one usually refers to the continuum of ZRS excitations as the upper Hubbard band. Exciton formation of the $(d^9L_h)_i -d^{10}_j$ pairs has been considered as a possible low-energy elementary excitation in insulating copper-oxide planes \cite{Sim96,Zha98} or chains \cite{Jec03,Geb97,Gal97,Bar02}.

In analogy to semiconductors, excitons may form in narrow-band Mott-Hubbard (MH) insulators if the attraction between holes and electrons exceeds their kinetic energy. In copper oxides, the exciton size corresponds to the interatomic Cu-Cu distance, and hence a large binding energy of the order of an eV is expected, which is characteristic of a tightly bound, Frenkel-type exciton. In contrast to Frenkel excitons in semiconductors, however, electrons in the MH model cannot be localized on the same site because of the large on-site Coulomb interaction, $U$, and an intrinsic broadening of the exciton linewidth is expected as a result of exchange interactions \cite{Don71}. It has therefore been difficult to separate the
intrinsically broad and weak features due to MH excitons from the strong
absorption continuum due to incoherent {\it p-d} transitions \cite{Pag02,Kim08}. Although peak structures arising from excitons have been identified in frequency-derivative reflectivity spectra \cite{Kim08}, an accurate determination of the exciton spectral weight (SW) and a quantitative comparison with model calculations have thus far not been reported.

In this Letter, we report a comprehensive ellipsometric study of charge excitations in LiCuVO$_4$, whose electronically active units are chains built up of edge-sharing CuO$_4$ plaquettes. In this lattice geometry, the Cu-O-Cu bonds form $\sim 90^\circ$ angles, such that the nearest-neighbor ($nn$) hopping amplitude is unusually weak and smaller than that between next-nearest neighbors ($nnn$). The resulting magnetic frustration and multiferroic properties have recently drawn much attention to this family of compounds \cite{Toh04,Dre07,Gib04,Sch08,Park07}. In our optical experiment we observe a weak but well-resolved two-peak structure for photon polarization along the chains. Using ellipsometry, we accurately determine its  temperature-dependent SW and show that it is controlled by the thermal variation of the $nn$ and $nnn$ spin correlation functions. The peaks can thus be identified as an exciton doublet arising from the long-range Coulomb interaction along the chains. By virtue of their exceptionally narrow electronic bandwidth, compounds with edge-sharing copper-oxide chains thus provide a highly favorable platform for the investigation of exciton formation and the interplay between spin and charge correlations in the cuprates.
In the CuO$_2$-planes of cuprate superconductors the band width is significantly larger, and hence excitons will not appear. However, a thorough understanding of the long-range Coulomb interaction is important for the quantitative description of features such as charge density waves and stripes.

LiCuVO$_{4}$ crystallizes in the orthorhombic space group \textit{Imma}, with copper-oxide chains running along the $b$-axis. Single crystals were grown from LiVO$_{3}$ flux, as described previously \cite{Pro00}. The magnetic susceptibility
exhibits two characteristic maxima at $T = 28$ K, due to the establishment of magnetic correlations within the Cu-O chains, and at 2.4 K, due to the formation of three-dimensional long-range order \cite{Gib04,End05,But07,Pro04}. For the optical measurements, the \(ab\) and \(ac\) surfaces
were polished to optical grade, using a 0.25 $\mu$m diamond suspension. The sample was mounted on the cold finger of a helium-flow cryostat with a base pressure of \(2 \times 10^{-9}\) Torr at room temperature. Ellipsometric measurements were performed with a  rotating-analyzer type Woollam VASE variable angle ellipsometer. Ellipsometry yields the anisotropic frequency-dependent complex dielectric tensor, $\tilde \varepsilon(\omega) = \varepsilon_1(\omega)+\emph{i}\varepsilon_2(\omega)=1+\emph{i}\ 4\pi\tilde \sigma(\omega)/\omega$, without the need for reference measurements
or Kramers-Kronig transformations.
A numerical regression procedure \cite{Elli} was applied to derive the principal components of the dielectric tensor from the ellipsometric data at different Euler angles.

Figure 1 shows the real and imaginary parts of the dielectric function
of LiCuVO$_{4}$ at 300, 100, and 7 K. In order to separate contributions from the different bands to the optical response, and to explore the origin of the temperature dependence, we performed a classical dispersion analysis with a minimum set of Lorentzian oscillators by simultaneous fitting to $\varepsilon_{1}(\omega)$ and $\varepsilon_{2}(\omega)$ \cite{EPAPS}. One can distinguish the two most prominent optical bands located at 3.7 eV (3.55 eV) [3.6 eV] and 4.2 eV (4.4 eV) [4.4 eV] in the spectra along the $a-$ ($b-$) [$c-$] axis. Clearly, the optical response is highly anisotropic. While the spectrum along the $c$-axis shows no significant temperature dependence, temperature dependent features are apparent along both $a$ and $b$. The main temperature dependent features (marked by arrows in Fig. 1) are centered at 4.2 eV for polarization along $a$ and at 2.95 eV along $b$.

A detailed analysis of the spectra reveals that the origin of the temperature dependence for both polarizations is qualitatively different. We first focus on polarization along $a$, perpendicular to the chains. The temperature-difference spectra $\Delta \sigma_1(\omega,T)$ and $\Delta \varepsilon_1(\omega,T)$ (with consecutive intervals $\Delta T =50$ K) displayed in Fig. 2(a) can be described mainly as a gradual narrowing and shift of the  optical band at 4.2 eV with decreasing temperature, without any discernible change in its intensity.
This behavior is typical for interband transitions and can be attributed to lattice anharmonicity. This conclusion is supported by the smooth temperature evolution of the $\sigma^a_1$ amplitude at $4.2$ eV and of $\varepsilon^{a}_{1}$  at $3.8$ and $4.45$ eV displayed in Fig. 2(c). The only break in this gradual evolution is a subtle kink in the temperature dependence of $\sigma^{a}_{1}$ and $\varepsilon^{a}_{1}$ at $T \sim 30$ K [Fig. 2(c)], close to the maximum of the magnetic susceptibility. This may indicate some influence of spin correlations on the inter-chain charge transfer excitations.

In contrast, the $\Delta\sigma^b_1(\omega)$ difference spectra along the chains [Fig. 2(b)] show the emergence of a well-defined absorption peak at 2.95 eV at low temperatures, which is accompanied by an antiresonance feature with zero-crossing
at the same energy in $\Delta\varepsilon^b_1(\omega)$. Figure 2(d) shows the temperature dependence of the $\sigma^b_1$ amplitude near the center of this peak (upper trace) and the corresponding changes in $\varepsilon^{b}_{1}$ measured at the off-resonant photon energies 2.65 and 3.45 eV (lower traces). These changes are Kramers-Kronig consistent and indicate a pronounced intensity enhancement of the band at 2.95 eV below 80 K, the temperature below which the magnetic susceptibility begins to deviate from the mean-field Curie-Weiss behavior due to the appearance of short-range
spin correlations along the copper-oxide chains \cite{Pro04}.

\begin{figure}[ht]
\includegraphics[width=7.0cm]{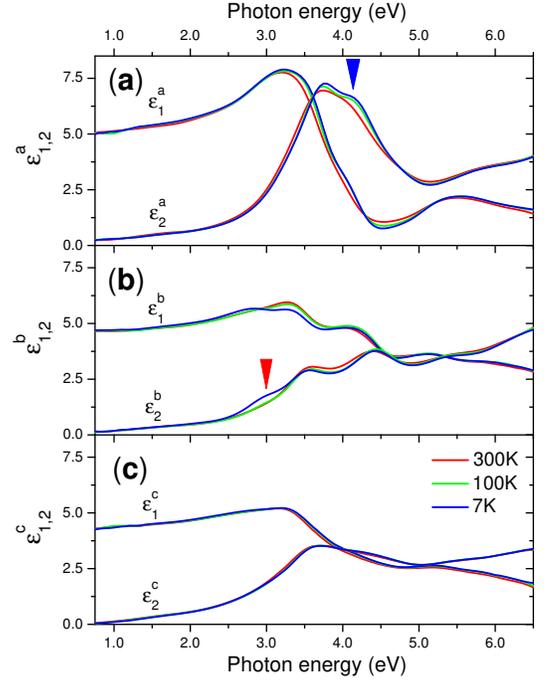}
\caption{Real $\varepsilon_1(\omega)$ and imaginary $\varepsilon_2(\omega) $  parts of the dielectric
function of LiCuVO$_{4}$  measured at $7$, $100$,
and $300$ K for photon polarizations (a) E \textbardbl\ $a$, (b) E \textbardbl\ $b$ and (c) E \textbardbl\ $c$. Blue and red arrows mark the 
temperature dependent features discussed in the text.}
\protect\label{fig1}
\end{figure}

The existence of positive and negative regions 
in the $\Delta\sigma^{a,b}_{1}(\omega)$  spectra [Fig. 2(a) and 2(b)] indicate a redistribution of optical SW. This can be quantified by integrating the optical conductivity in terms of the effective charge density $\Delta N_{eff}(\omega,T)=\frac{2m}{\pi e^2 N_{Cu}} \int_0^\omega \Delta\sigma_1(\omega^{'},T) d\omega^{'}$, where $m$ is the free electron mass and $N_{Cu}= 1.4\times 10^{22}$cm$^{-3}$
is the density of Cu atoms.
Figure 3(a) summarizes the low-temperature
changes in SW.  For polarization along the chains, the integral increases at low $\omega$ because of the band at $2.95$ eV,
but this gain is compensated by a SW loss 
within the spectral range of the bands at \(3.5\) and \(4.4\) eV, such that the optical sum rule is satisfied.
For polarization perpendicular to the chains, the  integral of $\Delta\sigma^{a}_{1}(\omega)$
is consistent with a  narrowing of the band at 4.2 eV upon cooling, and no SW redistribution between the \(3.7\) and \(4.2\) eV bands is indicated.

\begin{figure*}[ht]
\includegraphics[width=14.5cm]{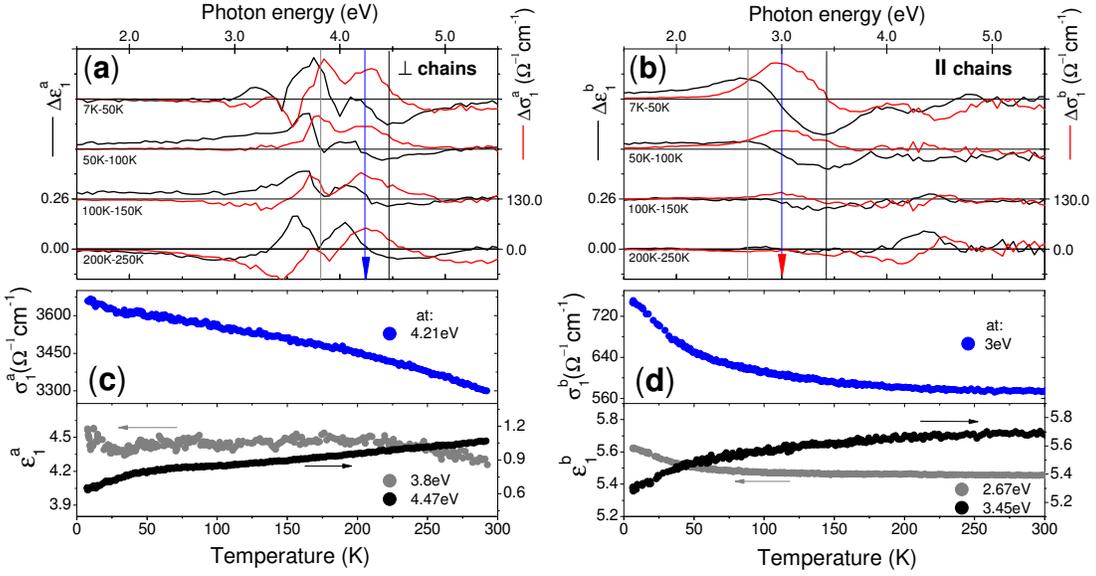}
\caption{(a), (b): Temperature-difference spectra $\Delta\sigma_1(T,\omega)=\sigma_1(T_1,\omega)-\sigma_1(T_2,\omega)$
(red) and $\Delta\varepsilon_1(T,\omega)=\varepsilon_1(T_1,\omega)-\varepsilon_1(T_2,\omega)$
(black) of LiCuVO$_{4}$ for polarizations (a) perpendicular to and (b) along the chains. Successive $\Delta\sigma_1(T,\omega)$ and $\Delta\varepsilon_1(T,\omega)$
spectra are shifted by 130  $\Omega^{-1}$cm$^{-1}$ and 0.26 for clarity \cite{EPAPS}.
The arrows mark the same energies as in Figs. 1(a) and 1(b). (c), (d):  Temperature dependence of $\sigma_1$ and $\varepsilon_1$ measured at $3.8$, $4.21$ and $4.45$ eV for polarization perpendicular to the chains and at $2.65$, $3.0$ and $3.45$ eV for polarization along the chains, as marked by vertical lines in (a) and (b), respectively. Cooling-down and warming-up curves are consistent and were averaged.}
\protect\label{fig2}
\end{figure*}

We now compare the salient observations to the results of model calculations, beginning with the overall anisotropy of the optical spectrum (Figs. 1 and
2). Since the main transitions are expected to be due to O$p^6$ Cu$d^9 \rightarrow$ O$p^5$ Cu$d^{10}$ excitations, we have computed the optical spectrum of the previously proposed \cite{Vernay08} $pd$-Hamiltonian, with on-site energies $\epsilon_p-\epsilon_d=3.5$ eV
and hopping parameters $t_{pd}=1.4$, $t^a_{pp}=0.55,t^b_{pp}=0.5, t^{2b}_{pp}=0.4$ eV [Fig. 3(b)]. While the overall agreement with the experimental data is reasonable, the mode at 2.95 eV is not described by the single-electron $pd$ model, and the strong temperature dependence of its SW suggests that it is generated by many-body effects. Recent calculations of the optical conductivity of edge-sharing copper-oxide chains within a $pd$ model including the on-site Coulomb repulsion \cite{Mal08} have assigned the lowest-energy
transition along the chains to ZRS excitations. A strong temperature dependence of the SW due to spin-correlation effects was also predicted, in qualitative agreement with our data.

\begin{figure}[ht]
\includegraphics[width=6.2cm]{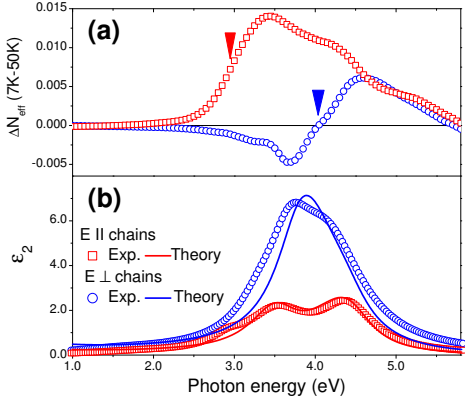}
\caption{(a) SW changes $\Delta N_{eff}(\omega)=N_{eff}(\omega,50$
K$)-N_{eff}(\omega,7  $ K$)$ in LiCuVO$_{4}$ for polarizations along (open
squares) and perpendicular (open circles)  to the chains. The arrows mark the same energies as in Figs. 1(a) and 1(b). (b) \(\varepsilon_{2}\) spectrum for polarizations along and perpendicular to the chains derived from experiment at \(7\) K and calculated from the $pd$-Hamiltonian as described in text. The optical bands above \(5\) eV listed in Table 1 were subtracted.
 }
\protect\label{fig3}
\end{figure}

However, close inspection of the ellipsometric data indicates a more complex behavior that cannot be understood in terms of incoherent ZRS excitations. Fig. 4(b) shows that the 2.95 eV band exhibits a double-peak structure with a satellite peak at \(2.15\) eV that is not visible in Figs.(1)-(3) because of its low intensity \cite{EPAPS}. The temperature dependence of the individual band SW was parameterized as  $\Delta N^{(l)}_{eff}(T)=\frac{2m}{N_{Cu}\pi e^2}\Delta SW^{(l)}(T)$ in Fig.4(c), where $\Delta SW^{(l)}(T)$ is determined by the oscillator strength of the individual band.
The SW of both bands increase upon cooling down to $\sim 80$ K in a parallel manner. For lower temperatures, however, the intensity of the lower-energy band saturates, while that of the higher-energy band keeps increasing.

\begin{figure}[ht]
\includegraphics[width = 7.0cm]{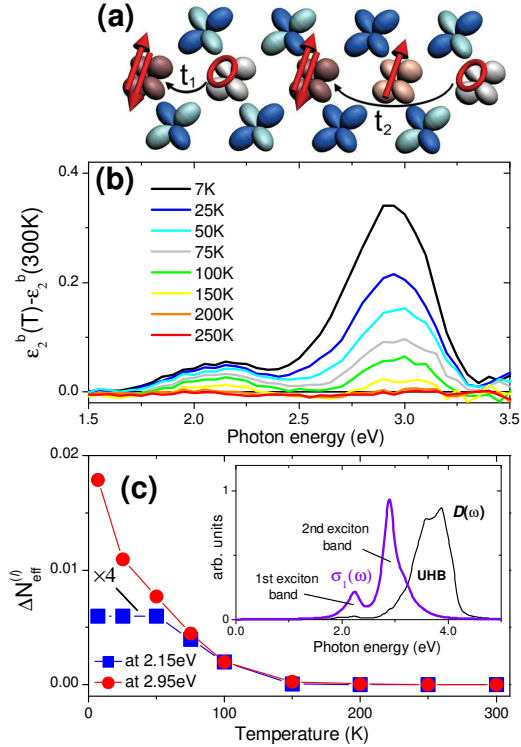}
\caption{ (a) Mott-Hubbard exciton formed by $d^9L_h$ and  $d^{10}$ states
generated by $nn$ ($t_1$) or $nnn$($t_2$) hopping. (b) Temperature-difference spectra of $\varepsilon_2(\omega)$
near \(2.9\) eV for polarization along the chains, background corrected due
to the temperature variation of the high-energy optical bands \cite{EPAPS}.
(c) Temperature dependence of SW of exciton bands at 2.15 and 2.95 eV. Inset: $\sigma_1(\omega)$ and density of states $D(\omega)$ of the UHB calculated with Eq. (1)  at $T=0$ and parameters $U=3.75$,
$V=1.6$, $t_1=0.08$ and $t_2=0.1$ eV. 
 }
\protect\label{fig4}
\end{figure}

We now demonstrate that the two bands can be quantitatively described as
an exciton doublet in a model incorporating long-range Coulomb interactions. We consider the one-band Hubbard model obtained
by downfolding from the \textit{pd}-Hamiltonian \cite{Oga08}. In the case of edge-sharing chains, it is important to consider both $nn$ $(t_1$) and $nnn$  ($t_2$) hopping [Fig. 4(a)]. Besides the kinetic energy terms
$K_{i,l}=-\sum_{\sigma} t_l(c^+_{i+l,\sigma}c_{i,\sigma}
+c^+_{i,\sigma}c_{i+l,\sigma}),$
the local Hubbard interaction $U$ and
the long-range Coulomb interaction $V_l$ are included. The model thus reads \cite{Hor05}:
\begin{equation}
H= \sum_{l=1}^2\sum_i K_{i,l}
+U\sum_{i} n_{i,\uparrow} n_{i,\downarrow}
+\sum_{i, l\geq 1} V_{l} n_i n_{i+l}  .
\end{equation}
Here $V_l=V/l$ is parameterized by the $nn$ Coulomb interaction $V$.
An optical excitation induced by the current operators $j_l=-i e\sum_{i,\sigma} d_l t_l
(c^+_{i+l,\sigma}c_{i,\sigma}-h.c.)$ generates an empty site (ZRS) and a doubly
occupied site (doublon in the upper Hubbard band, UHB) one ($d_1=b$) or two ($d_2=2 b$) Cu sites apart [Fig. 4(a)] \cite{Jec03}.
As a result of the Coulomb attraction 
between the positive hole and the negative doublon, exciton states emerge below the
UHB. For $nn$ hopping only, the lowest-energy exciton dispersion is small, $\sim - (2 t_1 \cos(k/2))^2/V$ \cite{Geb97, Gal97, Bar02}. In our case, however, the dispersion is large and given by
$E_1(k)=U-V\pm 2 t_2 \cos(k)$ due to the $t_2$ process, and the twofold degeneracy is lifted.
The second exciton is centered at $U-V/2$. As seen in the inset of Fig. 4(c), the SW is concentrated in the exciton states, and only little optical weight is found in the UHB, despite its large density of states. While this representation gives a correct description of the energy scales, one would expect sharp excitons if one considered the charge sector only. We argue, however, that
the coupling of the spin degrees of freedom (mainly due to $t_2$-processes)
implies a momentum average over the exciton band. This leads to a band with a width of $\sim 4 t_2 \sim 0.4$ eV for the lowest exciton, consistent with our experimental data.

The SW for the individual exciton transitions can be expressed in terms of
the kinetic energy per bond and the hopping length $d_l$ as
$\Delta N_{eff}^{(l)}=-\frac{m}{\hbar^2} d^2_l \langle K_{i,l}\rangle$.
The kinetic energy of the model given in Eq. (1) is equivalent to the superexchange energy of the corresponding Heisenberg model \cite{note}
$H_s=\sum_l J_l \sum_i(\vec{S_i}\cdot \vec{S}_{i+l}-1/4)-
J_1^F \sum_i \vec{S_i}\cdot \vec{S}_{i+l}$ where $J_l\simeq 4 t_l^2/(U-V_l)$.
Thus, the weights of the first and second excitons are directly related to the corresponding spin correlations:
\begin{equation}
\Delta N_{eff}^{(l)}=-\frac{2m}{\hbar^2} d^2_l J_l
\langle \vec{S_i}\cdot \vec{S}_{i+l}-1/4 \rangle .
\end{equation}
This relation explains why the second exciton is much stronger: (i) $d_2=2 d_1$ and (ii) the $nn$ spin-correlation function ($l=1$) is frustrated and small, because  $J_1-J_1^F$ is ferromagnetic and small compared to the antiferromagnetic coupling constant $J_2$. For $l=2$, the correlation function is large and negative, leading to large variation in $\Delta N_{eff}^{(2)}$ as a function of temperature. These arguments also explain why the first exciton band experimentally observed at 2.15 eV is suppressed below 80 K, whereas the second band at 2.95 eV shows a steep increase [Fig. 4(c)] following the spin-correlation effects. A numerical computation of Eq. (2) using exchange parameters determined by neutron scattering \cite{End05} estimates within a factor of two the experimentally determined $\Delta N_{eff}^{(2)}(75$ K$-7$ K$)\simeq
0.013$. Within the model the exciton mode draws its SW from the UHB, which may be superimposed on the charge-transfer excitations at 3.5 and 4.4 eV [Fig. 3(a)].

In summary, our optical measurements have yielded clear evidence for exciton formation and interesting insights into the relationship between the charge dynamics and frustrated magnetism in a Mott-Hubbard insulator with edge-sharing copper-oxide chains. The results and methodology established here are a good starting point for further investigations of compounds with doped copper-oxide chains and planes.

\bibliographystyle{apsrev}

\begin{thebibliography}{99}
\bibitem{note}
The ferromagnetic coupling $J_1^F$ originates from a
two hole state on O that is not included in Eq. (1), but
must be considered for the magnetism.
\end{thebibliography}


\newpage

\widetext

\begin{center}\begin{Large}\textbf{EPAPS supplementary online material: 
\\{"Exciton doublet in the Mott-Hubbard LiCuVO$_4$ insulator\\ 
identified by spectral ellipsometry"}}
\end{Large}\end{center}

\begin{large}

\begin{itemize}
\item\ 
\textbf{Anisotropic dielectric response of LiCuVO$_{4}$: overall features and dispersion analysis}
\end{itemize}

\renewcommand{\thefigure}{S1}

\begin{figure}[ht]
\includegraphics[width=14cm]{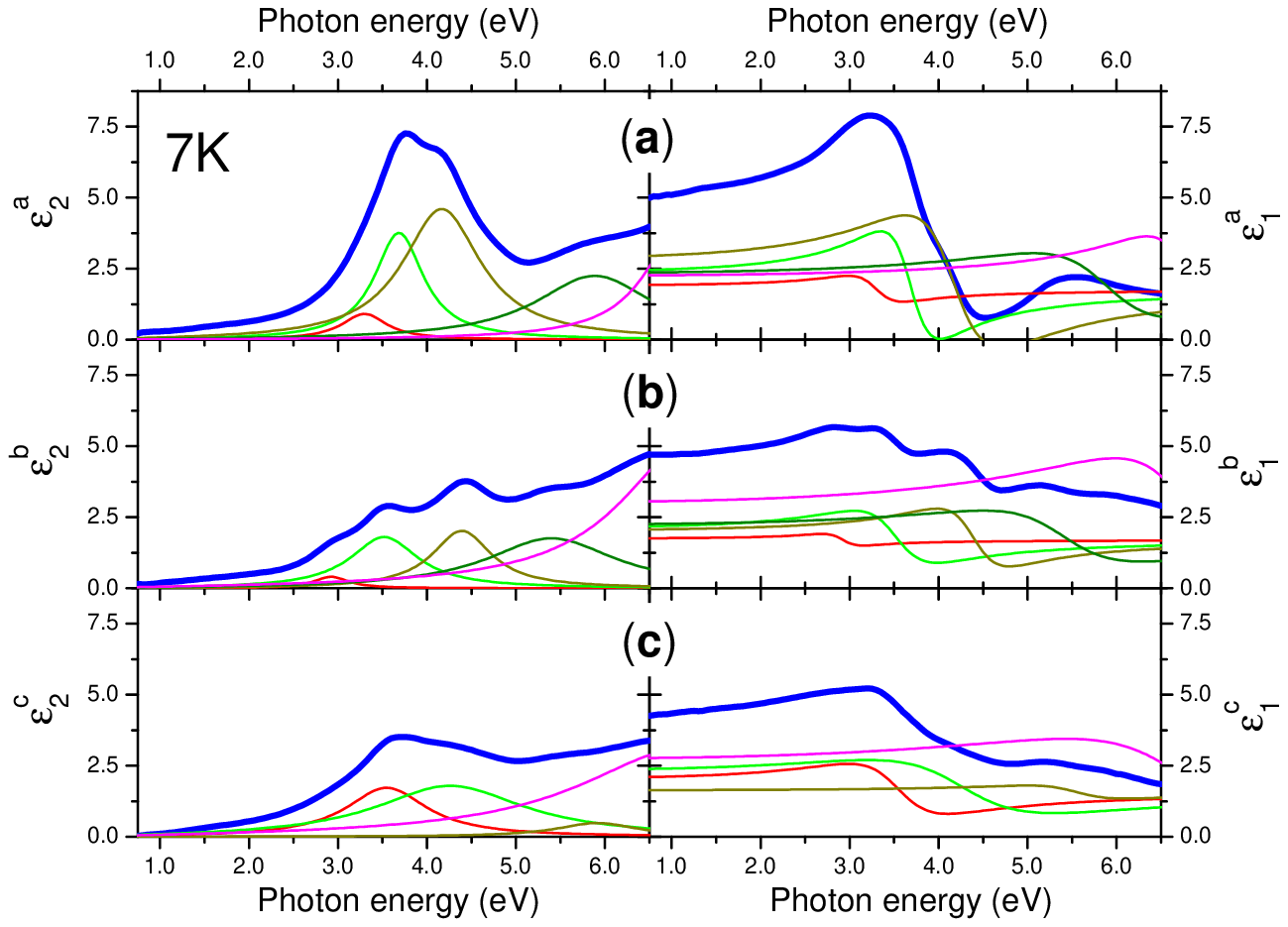}
\floatname{type}{floatname}
\caption{Real $\varepsilon_1(\omega) $ and imaginary $\varepsilon_2(\omega)$ parts of the dielectric function of LiCuVO$_{4}$ measured at $7$K for photon polarizations (a) E \textbardbl\ $a$, (b) E \textbardbl\ $b$ and (c) E \textbardbl\ $c$ and represented by the individual contributions of separate Lorentzian bands. The parameters are listed in Table S1. }

\protect\label{fig1}
\end{figure}

In order to separate contributions from the different bands to the optical response of LiCuVO$_{4}$, we performed a classical dispersion analysis by simultaneous fitting to $\varepsilon_{1}(\omega)$ and $\varepsilon_{2}(\omega)$. The decomposition of the optical response in terms of a sum of Lorentzians is a phenomenological fitting procedure to quantitatively parameterize the observed temperature-driven optical anomalies and spectral weight transfer.

Moreover, the agreement between the simulated and measured spectra confirms that our data for $\epsilon_1(\omega )$ and $\epsilon_2(\omega )$ (or $\sigma_1(\omega )$) determined independently by spectroscopic ellipsometry are Kramers-Kronig consistent.

\renewcommand{\thetable}{S1}

\begin{table}[b]
\caption{\label{tab:table2}Parameters of Lorentz oscillators resulting from a dispersion analysis of complex dielectric response in $a$-axis ($b$-axis)
[$c$-axis] polarization in LiCuVO\(_{4}\) measured at \(T=7\) K. \(\varepsilon_{\infty}\)=1.75
(\(\varepsilon_{\infty}\)=1.69) [\(\varepsilon_{\infty}\)=1.55].  }
\begin{ruledtabular}
\begin{tabular}{lll}
$\omega_j$(eV)&$S_j$ (eV\(^{2}\))&$\Gamma_j$(eV)\\
\hline
3.31 (2.93)& 1.86 (0.52) & 0.62 (0.47)\\
3.70 (3.55) [3.59] & 9.30 (5.92) [6.83] & 0.67 (0.93) [1.11]\\
4.20 (4.41) [4.38] & 20.6 (7.19) [15.7] & 1.07 (0.81) [2.03]\\
5.94 (5.47) [5.58] & 21.5 (16.8) [2.80] & 1.62 (1.76) [1.11]\\
6.86 (7.00)  [7.00] & 24.0 (66.2) [59.3] & 1.00 (1.90) [2.79]\\
\end{tabular}
\end{ruledtabular}
\end{table}

A minimum set of Lorentzian oscillators, with one high-energy oscillator beyond the investigated spectral range, was introduced to represent a dielectric function in the form $\tilde\varepsilon(\omega) =\varepsilon_{\infty}+ \sum_j\frac{S_j}{\omega_j^2-\omega^2-i\omega\Gamma_j}$, where $\omega_j$, $\Gamma_j$, and $S_j$ are the peak energy, width, and 
oscillator strength of the $j$th oscillator, and $\varepsilon_{\infty}$ is the core contribution from the dielectric function. The parameters determined
by simultaneous fitting to $\varepsilon_{1}(\omega)$ and $\varepsilon_{2}(\omega)$
measured at 7 K are listed in Table S1.

\begin{itemize}
\item
\textbf{Temperature-difference spectra}
\end{itemize}

Figure S2 shows the temperature-difference spectra
$\Delta\sigma_1(T,\omega)=\sigma_1(T_1,\omega)-\sigma_1(T_2,\omega)$
and $\Delta\varepsilon_1(T,\omega)=\varepsilon_1(T_1,\omega)-\varepsilon_1(T_2,\omega)$
with consecutive intervals $\Delta T =50$ K and represents
an expanded version of Figs. 2(a) and 2(b) of the manuscript. 

\renewcommand{\thefigure}{S2}

\begin{figure}[ht]
\includegraphics[width=18cm]{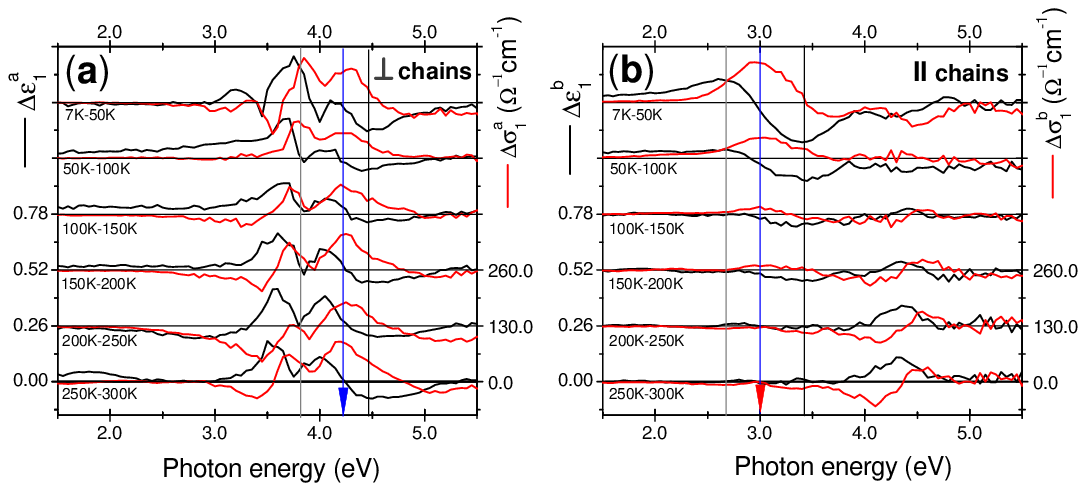}
\caption{(a), (b): Temperature-difference spectra $\Delta\sigma_1(T,\omega)=\sigma_1(T_1,\omega)-\sigma_1(T_2,\omega)$
(red) and $\Delta\varepsilon_1(T,\omega)=\varepsilon_1(T_1,\omega)-\varepsilon_1(T_2,\omega)$
(black) of LiCuVO$_{4}$ for polarizations (a) perpendicular to and (b) along the chains. Successive $\Delta\sigma_1(T,\omega)$ and $\Delta\varepsilon_1(T,\omega)$
spectra are shifted by 130  $\Omega^{-1}$cm$^{-1}$ and 0.26 for clarity.
The arrows mark the same energies (4.2 eV and 2.95 eV, respectively) as in Figs. 1(a) and 1(b) of the manuscript.}
\protect\label{fig1}
\end{figure}

\begin{itemize}
\item
\textbf{Optical response along the chains: Double-peak structure near 2.9 eV}
\end{itemize}

\begin{figure}[ht]
\begin{minipage}[b]{0.49\linewidth}
\centering
\includegraphics[width=10cm]{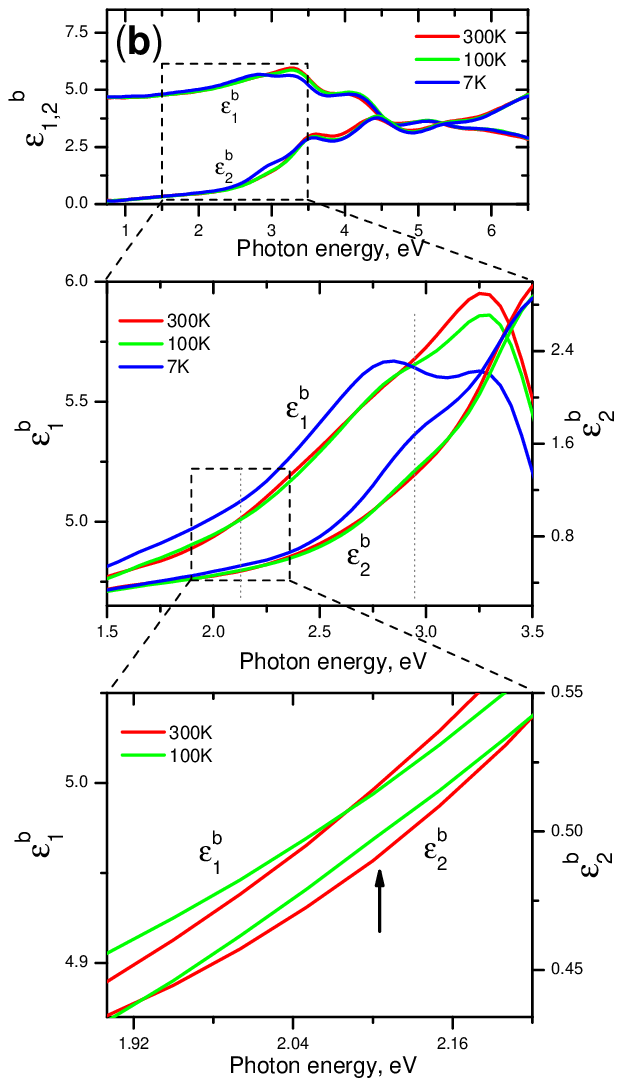}
\renewcommand{\thefigure}{S3}
\caption{Real $\varepsilon_1(\omega)$ and imaginary $\varepsilon_2(\omega)$
parts of the dielectric function of LiCuVO$_{4}$ measured at $7$, $100$,
and $300$ K in $b-$axis polarization. Upper panel is the same as in Fig.
1(b) of the manuscript. Lower panels show an enlarged view of the upper panel. }
\protect\label{fig1}
\end{minipage}
\hspace{0.2cm}
\begin{minipage}[b]{0.49\linewidth}
\centering
\includegraphics[width=10cm]{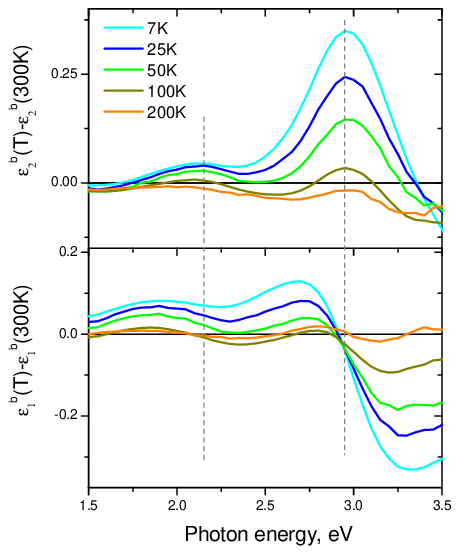}
\renewcommand{\thefigure}{S4}
\caption{Temperature-difference spectra of imaginary $\varepsilon_2(\omega)$
(upper panel) and real $\varepsilon_1(\omega)$ (lower panel) parts of the dielectric function of LiCuVO$_{4}$ for polarization along the chains. Vertical
dashed lines mark photon energies 2.15 eV and 2.95 eV at which resonant behavior
in  $\Delta \varepsilon_2(\omega)$ coincides with zero crossing in $\Delta \varepsilon_1(\omega)$.}
\protect\label{fig1}
\end{minipage}
\end{figure} 
Figure S3  shows  that the emergence of the satellite absorption peak at 2.15 eV at low T is evident from the obvious changes in both the imaginary,
$\varepsilon_2(\omega)$, and real, $\varepsilon_1(\omega)$,  parts of the complex dielectric function presented in Fig.1(b) of the manuscript. The arrow in the lower panel marks, in particular, the photon energy at which the positive changes in $\varepsilon_2(\omega)$ coincides with crossing in
$\varepsilon_1(\omega)$, as temperature decreases from 300 K to 100 K.   These features are clearly resolved in the temperature-difference spectra shown in Fig.S4, as a "bump" in $\Delta\varepsilon_2(\omega)$ and a "wiggle"
in $\Delta \varepsilon_1(\omega)$, and can be quantitatively parameterized
by the Lorentzian fit. From this fit, the temperature dependence of the individual
band spectral weight is determined by the oscillator strength of this band,
$\Delta SW^{(l)}(T)=S_l(T)[eV^2]$. Alternatively, one can integrate the real
part of the optical conductivity within the individual band, $\Delta SW^{(l)}(T)=1/4\pi\int^{(l)}{\omega\Delta\varepsilon_2(\omega)d\omega}$,
after subtraction of the electronic background due to the temperature variation of the high-energy
optical bands. The gradual changes of the electronic background with temperature
are well controlled by the dispersion analysis of the whole $\varepsilon_1(\omega)$ and $\varepsilon_2(\omega)$ spectra performed at every temperature. Figure
4(b) of the manuscript shows the background corrected temperature differences
spectra of $\varepsilon_2(\omega)$ near 2.9 eV. Both the above procedures
give the same temperature dependence of the individual band spectral weight, the uncertainty is within the symbol size in Fig.4(c) of the manuscript.

\end{large}

\end{document}